\documentclass[]{spie}

\usepackage{graphicx}
\usepackage{subcaption} 

\graphicspath{{images/}}

\title{The TP3-WFS: a new guy in town}

\author[a,b]{Carlos Colodro-Conde}
\author[a,b]{Sergio Velasco}
\author[a,b]{Roberto L\'opez}
\author[a,b]{Alejandro Oscoz}
\author[a,b]{Yolanda Mart\'in-Hernando}
\author[a,b,c]{Rafael Rebolo}
\author[d]{Antonio P\'erez-Garrido}
\author[h,i,j]{Juan Jos\'e Ferr\'andez Valdivia}
\author[e]{Lucas Labadie}
\author[f]{Craig Mackay}
\author[a,b]{Marta Puga}
\author[g]{Gustavo Rodr\'iguez-Coira}
\author[a,b]{Luis Fernando Rodr\'iguez-Ramos}
\author[h,i,j]{Jos\'e Manuel Rodr\'iguez-Ramos}

\affil[a]{Instituto de Astrof\'isica de Canarias, Spain}
\affil[b]{Departamento de Astrof\'isica, Universidad de La Laguna, Spain}
\affil[c]{Consejo Superior de Investigaciones Cient\'ificas, Spain}
\affil[d]{Departamento de F\'isica Aplicada, Universidad Polit\'ecnica de Cartagena, Spain}
\affil[e]{I. Physikalsiches Institut, Universit\"{a}t zu K\"{o}ln, Germany}
\affil[f]{Institute of Astronomy, University of Cambridge, United Kingdom}
\affil[g]{Observatoire de Paris, France}
\affil[h]{Departamento de Ingenieria Industrial, Universidad de La Laguna, Spain}
\affil[i]{Wooptix S.L., Spain}
\affil[j]{Centro de Investigaciones Biom\'edicas de Canarias, Spain}

\begin{document} 
\maketitle

\begin{abstract}
The TP3-WFS (Two Pupil Plane Positions Wavefront Sensor) is, to the best of the authors' knowledge, the first physical implementation of the geometrical wavefront reconstruction algorithm that has been tested in a telescope as part of an actual AO-enabled instrument. The main advantage of this algorithm is that it theoretically provides a fairly good reconstruction accuracy even under very low levels of light, in the order of a few hundreds of photons. This paper presents the first control-related results obtained at the William Herschel Telescope (WHT). Such results demonstrate the feasibility of this novel WFS as part of a real-time AO control loop.
\end{abstract}


\section{INTRODUCTION}
\label{sec:intro}

The TP3-WFS was tested as part of the AOLI (Adaptive Optics Lucky Imager) instrument \cite{2017hsa9.conf..696V, doi:10.1093/mnras/stx262} during two short commissioning runs at the 4.2m William Herschel Telescope (WHT) in Roque de los Muchachos observatory (La Palma, Spain) during 2017. AOLI combines two well-known high-resolution techniques that had never been specifically combined in an astronomical instrument before: Adaptive Optics (AO) and Lucky Imaging (LI).

The geometrical wavefront reconstruction algorithm implemented in the TP3-WFS, initially proposed by Marcos Van Dam \cite{vanDam:02}, requires two defocused pupil planes: one at each side at the actual pupil plane. Unlike curvature wavefront sensors, this algorithm does not calculate the second derivative of the wavefront. Instead, it assumes a purely geometrical propagation of the wavefront, and calculates the first order derivatives directly. Such assumption is only valid with relatively short defocus distances, where diffraction effects do not have a relevant influence on the produced images.

\begin{figure}
	\centering
	\includegraphics[scale=0.4]{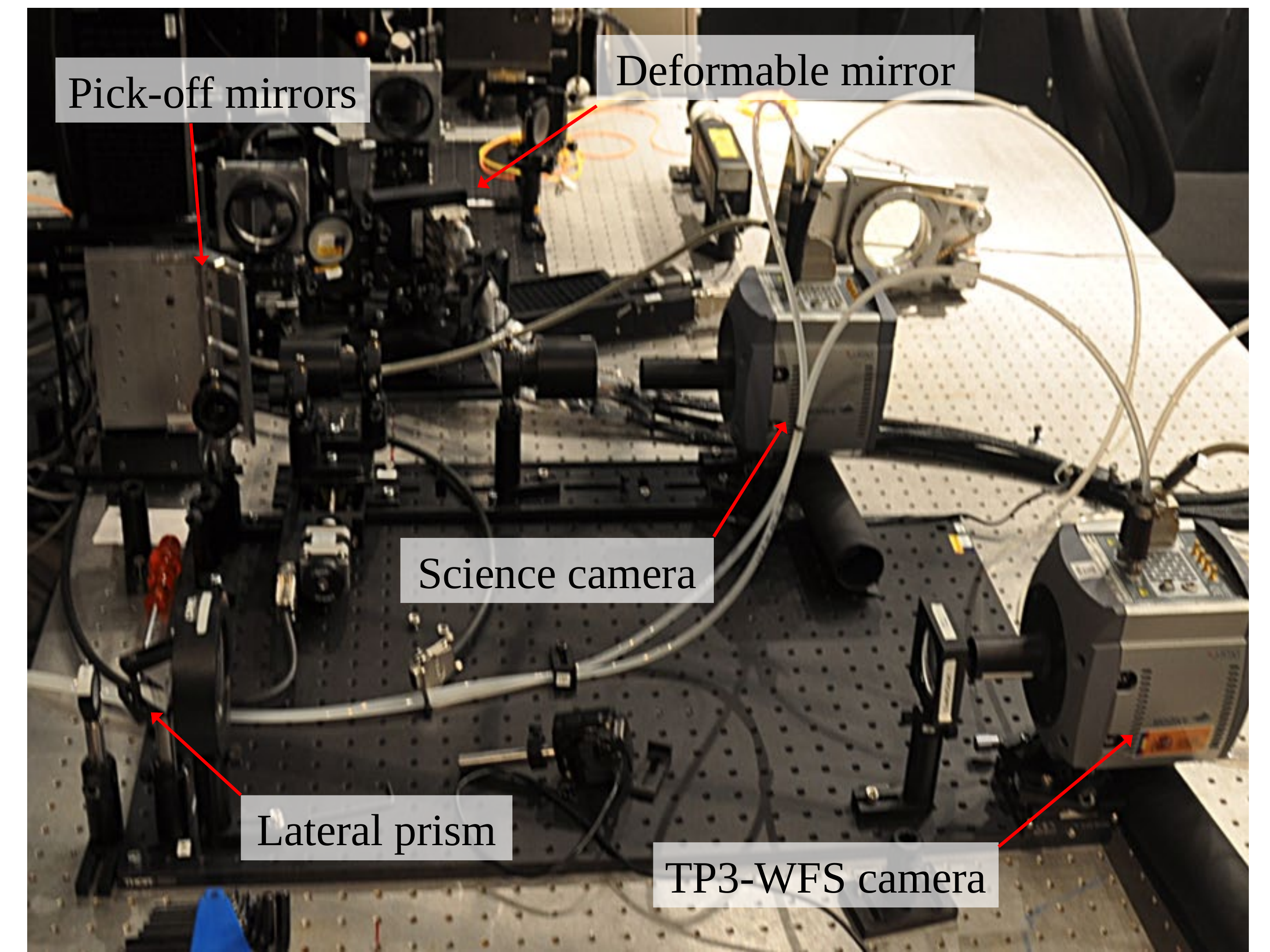}
	\caption {AOLI mounted at the WHT, showing the different subsystems. The defocused images are produced by means of a lateral prism which	splits the input beam in two, resulting in two optical paths with different	lengths. The two defocused images are then acquired by a single camera.}
	\label{fig:aoli_photo_labels}
\end{figure}

The input of the AO control system was our TP3-WFS, and the output was a 241-actuator deformable mirror (DM) by ALPAO. The TP3-WFS operates with the images provided by an Andor Ixon DU-897 camera, which is based on a sub-photon noise 512x512 e2v EMCCD (Electron Multiplying Charge-Coupled Device) detector. The defocused pupil images that are required by the wavefront reconstruction algorithm were produced by means of a lateral prism which splits the input beam in two, resulting in two optical paths with different lengths. The two defocused images are then acquired by a single camera. A photograph of the instrument as mounted at the WHT is shown in Figure~\ref{fig:aoli_photo_labels}, including labels that highlight the main elements that comprise the AO system.


\section{SOFTWARE}

The geometric reconstruction algorithm runs in an nVidia GeForce GTX Titan Z GPU, by means of the CUDA language. Among all the possible acceleration methods, it was decided that the wavefront reconstruction algorithm would take advantage of GPUs in order to achieve real-time operation. More specifically, the calculation of derivatives, diffractions, interpolations, rotations, and matrix products are accelerated with the aid of the CUDA language by means of custom GPU kernels. Other simple pre-processing operations like image cropping and bias subtraction are executed on the CPU. Our current implementation is able to reconstruct 153 Zernike modes with a delay of approximately 1.1 ms.

\begin{figure}
	\centering
	\includegraphics[scale=0.9]{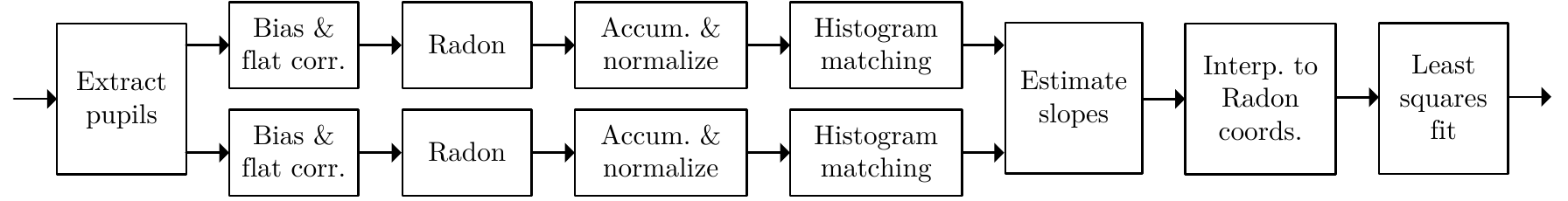}
	\caption {Steps of the geometrical wavefront reconstruction algorithm. The Radon transform estimates the amount of displacement suffered by the photons while they travel from one defocused plane to the other. The slopes of the wavefront can be easily calculated from such displacements if one assumes geometrical optics. A least squares fit can be then applied to obtain a Zernike representation of the wavefront.}
	\label{fig:wfr_arch}
\end{figure}

Figure~\ref{fig:wfr_arch} summarizes the steps that define the algorithm. The step that clearly has a higher computational cost is the calculation of the Radon transform \cite{Radon17}, which is used to estimate the amount of displacement suffered by the photons while they travel from one defocused plane to the other. The slopes of the wavefront can be easily calculated from such displacements if one assumes geometrical optics, following the steps defined in \cite{vanDam:02}. A least squares fit can be then applied to obtain a Zernike representation of the wavefront.

The TP3-WFS software possesses a handy GUI (Graphical User Interface) that eases the configuration of the wavefront reconstruction algorithm. Among other parameters, the GUI allows defining the two regions of the input image that correspond to each pupil image, the number of projection angles for which the Radon transform will be calculated and the number of reconstructed Zernike modes. The relationship between these last two parameters is of great importance, as there is a minimum number of angles that need to be calculated so as to ensure that the least-squares fit is being executed correctly for a given number of modes.

\section{RESULTS}

In the context of our tests, the TP3-WFS was configured to reconstruct 153 Zernike modes using 31 projection angles. For this number of Zernike modes, more projection angles would have increased the processing time with no benefit in the quality of the reconstructed wavefront. The radius of the defocused pupil images on the detector, determined by the parameters of the optics, was 30 pixels. For simplicity, only bias corrections were applied to the extracted images this time.

The TP3-WFS was first validated in the laboratory by measuring the influence matrix of the AO system. The same type of measurement was executed at the telescope using a natural reference star. In both cases, the inspection of the influence matrices demonstrated that the actuators had been properly detected. The elevated number of reconstructed modes made it possible for the actuator peaks to keep their expected shape- Nominal operations in AOLI would use a lower number of reconstructed modes, as the objective of the AO subsystem is to produce images with enough quality for the LI algorithm to be applicable.

\begin{figure}
	\centering
	\begin{subfigure}{0.24\textwidth}
	\centering
	\includegraphics[scale=0.8]{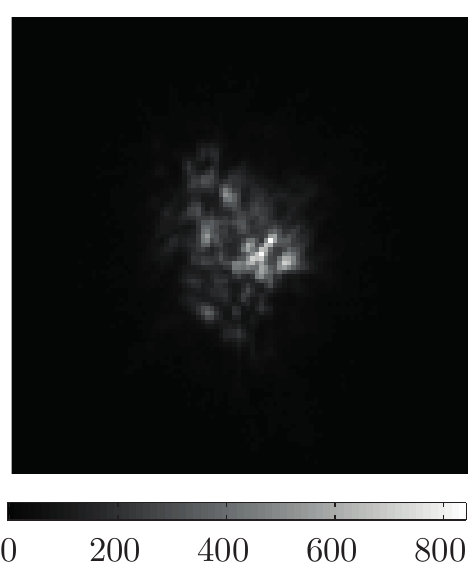}
	\caption{Open loop}
	\label{fig:typ_ol} 
	\end{subfigure}
	\begin{subfigure}{0.24\textwidth}
	\centering
	\includegraphics[scale=0.8]{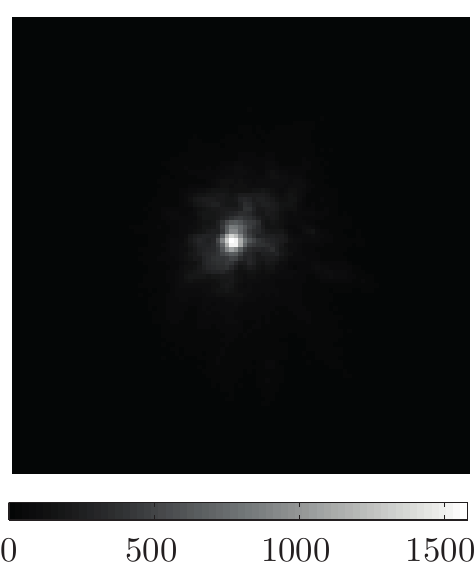}
	\caption{Closed loop}
	\label{fig:typ_cl}
	\end{subfigure}
	\caption {Typical speckle images taken by the science camera, with the AO loop disabled (a) and enabled (b). Only in the latter case it is possible to apply lucky imaging, which is just the purpose of the AOLI instrument.}
	\label{fig:science}
\end{figure}

The complete AO system was then tested in closed loop by pointing to a bright natural star. The activation of the control loop clearly reduces the wavefront error as seen by the TP3-WFS. What is more, the results as seen by the science camera (Figure~\ref{fig:science}) show that the AO corrections increase the probability of getting less-aberrated images, which may enable to apply the LI algorithm and thus accomplish the purpose of the AOLI instrument.

\section{FUTURE WORK}

Future implementations of the TP3-WFS optics will be based on mirrors so as to minimize chromatic dispersion at the sensor. This newer system will additionally allow configuring the defocus distances so as to study the effect of this parameter on dynamic range and sensibility.

Regarding the software, further effort will be put in optimizing the GPU implementation the wavefront reconstruction algorithm, so as to reduce the delay of the AO control loop. The design of this software will be revised to maximize parallelism and to take advantage of GPU textures to implement rotations and interpolations.

Finally, the algorithm will be tested under extremely low light conditions, that is, in the order of few hundreds of photons. The objective is to be able to perform AO corrections with faint natural stars, thus increasing the sky coverage of the instrument. Simulations in previous works \cite{vanDam:02} are promising but noise sources at the detector (e.g., clock induced charge) may prevent a clean detection of the photon events, which would have a negative impact on reconstruction accuracy.

\acknowledgments

This work was supported by the Spanish Ministry of Economy under the projects AYA2011-29024, ESP2014-56869-C2-2-P, ESP2015-69020-C2-2-R and DPI2015-66458-C2-2-R, by project 15345/PI/10 from the Fundaci\'on S\'eneca, by the Spanish Ministry of Education under the grant FPU12/05573, by project ST/K002368/1 from the Science and Technology Facilities Council and by ERDF funds from the European Commission. The results presented in this paper are based on observations made with the William Herschel Telescope operated on the island of La Palma by the Isaac Newton Group in the Spanish Observatorio del Roque de los Muchachos of the Instituto de Astrof\'isica de Canarias.

\bibliography{tp3_ao4elt5}

\begin{thebibliography}{1}

\bibitem{2017hsa9.conf..696V}
{Velasco}, S., {Colodro-Conde}, C., {L{\'o}pez}, R.~L., {Oscoz}, A.,
  {Valdivia}, J.~J.~F., {Rebolo}, R., {Femen{\'{\i}}a}, B., {King}, D.~L.,
  {Labadie}, L., {Mackay}, C., {Muthusubramanian}, B., {P{\'e}rez-Garrido}, A.,
  {Puga}, M., {Rodr{\'{\i}}guez-Coira}, G., {Rodr{\'{\i}}guez-Ramos}, L.~F.,
  and {Rodr{\'{\i}}guez-Ramos}, J.~M., ``{Completing the puzzle: AOLI
  full-commissioning fresh results and AIV innovations},'' in [{\em Highlights
  on Spanish Astrophysics IX}{\nolinebreak\hspace{0.1em}]},  {Arribas}, S.,
  {Alonso-Herrero}, A., {Figueras}, F., {Hern{\'a}ndez-Monteagudo}, C.,
  {S{\'a}nchez-Lavega}, A., and {P{\'e}rez-Hoyos}, S., eds.,  696--700 (Mar.
  2017).

\bibitem{doi:10.1093/mnras/stx262}
Colodro-Conde, C., Velasco, S., Fernández-Valdivia, J.~J., López, R., Oscoz,
  A., Rebolo, R., Femenía, B., King, D.~L., Labadie, L., Mackay, C.,
  Muthusubramanian, B., Pérez~Garrido, A., Puga, M., Rodríguez-Coira, G.,
  Rodríguez-Ramos, L.~F., Rodríguez-Ramos, J.~M., Toledo-Moreo, R., and
  Villó-Pérez, I., ``Laboratory and telescope demonstration of the tp3-wfs
  for the adaptive optics segment of aoli,'' {\em Monthly Notices of the Royal
  Astronomical Society}~{\bf 467}(3),  2855--2868 (2017).

\bibitem{vanDam:02}
van Dam, M.~A. and Lane, R.~G., ``Wave-front sensing from defocused images by
  use of wave-front slopes,'' {\em Appl. Opt.}~{\bf 41},  5497--5502 (Sep
  2002).

\bibitem{Radon17}
Radon, J., ``{\"U}ber die {B}estimmung von {F}unktionen durch ihre
  {I}ntegralwerte l\"angs gewisser {M}annigfaltigkeiten,'' {\em Berichte der
  Sachsischen Akadamie der Wissenschaft}~{\bf 69},  262--277 (1917).

\end{thebibliography}
\bibliographystyle{spiebib}

\end{document}